\documentclass{article}

\usepackage{arxiv}

\usepackage[utf8]{inputenc} 
\usepackage[T1]{fontenc}    
\usepackage{hyperref}       
\usepackage{url}            
\usepackage{booktabs}       
\usepackage{amsfonts}       
\usepackage{nicefrac}       
\usepackage{microtype}      
\usepackage{lipsum}
\usepackage{amsmath}
\usepackage{amssymb}
\usepackage{graphicx}
\usepackage{siunitx}
\usepackage{subfig}
\newcommand{\argmin}{\mathop{\mathrm{argmin}}}

\title{Learned Cone-Beam CT Reconstruction \\Using Neural Ordinary Differential Equations}

\author{
  Mareike Thies\\
  Pattern Recognition Lab\\ 
  Friedrich-Alexander Universit\"at\\
  Erlangen-N\"urnberg\\
  \texttt{mareike.thies@fau.de} \\
  \And
  Fabian Wagner\\
  Pattern Recognition Lab\\ 
  Friedrich-Alexander Universit\"at\\
  Erlangen-N\"urnberg\\
  \texttt{fabian.wagner@fau.de} \\
  \And
  Mingxuan Gu\\
  Pattern Recognition Lab\\ 
  Friedrich-Alexander Universit\"at\\
  Erlangen-N\"urnberg\\
  \texttt{mingxuan.gu@fau.de} \\
  \And
  Lukas Folle\\
  Pattern Recognition Lab\\ 
  Friedrich-Alexander Universit\"at\\
  Erlangen-N\"urnberg\\
  \texttt{lukas.folle@fau.de} \\
  \And
  Lina Felsner\\
  Pattern Recognition Lab\\ 
  Friedrich-Alexander Universit\"at\\
  Erlangen-N\"urnberg\\
  \texttt{lina.felsner@fau.de} \\
  \And
  Andreas Maier\\
  Pattern Recognition Lab\\ 
  Friedrich-Alexander Universit\"at\\
  Erlangen-N\"urnberg\\
  \texttt{andreas.maier@fau.de} \\
}

\begin{document}
\maketitle

\begin{abstract}
Learned iterative reconstruction algorithms for inverse problems offer the flexibility to combine analytical knowledge about the problem with modules learned from data. This way, they achieve high reconstruction performance while ensuring consistency with the measured data. In computed tomography, extending such approaches from 2D fan-beam to 3D cone-beam data is challenging due to the prohibitively high GPU memory that would be needed to train such models. This paper proposes to use neural ordinary differential equations to solve the reconstruction problem in a residual formulation via numerical integration. For training, there is no need to backpropagate through several unrolled network blocks nor through the internals of the solver. Instead, the gradients are obtained very memory-efficiently in the neural ODE setting allowing for training on a single consumer graphics card. The method is able to reduce the root mean squared error by over 30\% compared to the best performing classical iterative reconstruction algorithm and produces high quality cone-beam reconstructions even in a sparse view scenario.  
\end{abstract}


Extending analytical or iterative reconstruction algorithms for computed tomography (CT) by deep learning modules has shown to improve the quality of the reconstructed images, especially in challenging cases like high noise levels or insufficient projection data \cite{chen2018learn, jin2017}. For CT reconstruction, the input and output data of the problem are connected in a non-trivial geometrical manner. This is the reason why most deep-learning-based reconstruction approaches, instead of learning direct mapping from sinogram to image domain, incorporate knowledge about the physical operator connecting both domains and replace single components in the reconstruction pipeline by their learned counterpart, mostly operating in one domain only \cite{adler2021}. 

Unrolled iterative approaches seek to solve the reconstruction problem by loosely mimicking known iterative algorithms for inverse problems. This is done by unrolling a fixed number of iterations in depth as a deep learning architecture and incorporating learnable components in each step which are trained end-to-end. The exact architecture and the role of the trainable modules can vary giving rise to a number of approaches for MRI \cite{hammernik2018, gadjimuradov2021} and CT \cite{chen2018learn, vishnevski2019, adler2018}. 

While these unrolled iterative algorithms have achieved superior performance for the reconstruction of 2D images, their extension to the 3D case is challenging. Training requires gradient backpropagation through the entire unrolled sequence of trainable and known operators, thereby consuming a large amount of memory on the graphics card (GPU). In the 3D case, the amount of memory occupied by intermediate representations needed during backpropagation exceeds the memory of modern GPUs making the direct application of iterative approaches infeasible. 

Previous work addressed the prohibitively high memory consumption of unrolled 3D models. Greedy training of each iteration independently reduces memory consumption and allows training on patches but does not result in an optimal joint weight configuration \cite{wu2019}. When increasing the volume resolution with network depth, memory consumption is dominated by the single final iteration on full scale, but image quality is coupled to the expressiveness of the last iteration \cite{hauptmann2020}. Further, the use of invertible networks avoids storing intermediate representations which makes the memory requirement constant in depth but requires the network architecture to meet certain criteria for invertibility \cite{rudzusika2021, kellman2020}.      

In this work, we propose to interpret the series of unrolled iterations as a continuous residual process and formulate the problem in terms of an ordinary differential equation (ODE). This allows us to map the reconstruction problem onto an initial value problem which can be solved and trained memory-efficiently using recently proposed neural ODEs \cite{chen2018}. The key idea is that the memory requirement does not depend on the number of iterations, i.e., network depth. Instead, the forward pass is replaced by a call to an ODE solver and gradients are obtained by solving another adjoint ODE without storing the intermediate representations of the forward pass. Whereas a similar idea has been applied to MRI reconstruction \cite{chen2020}, to the best of our knowledge we are the first ones to apply neural ODEs to CT reconstruction. We show that using this method we obtain 3D cone-beam CT reconstructions from few angles with superior image quality compared to classical analytic and algebraic algorithms.

\section{Methods}

\begin{figure}[!t]
    \centering
    \subfloat[]{
        \includegraphics[height=2.9cm]{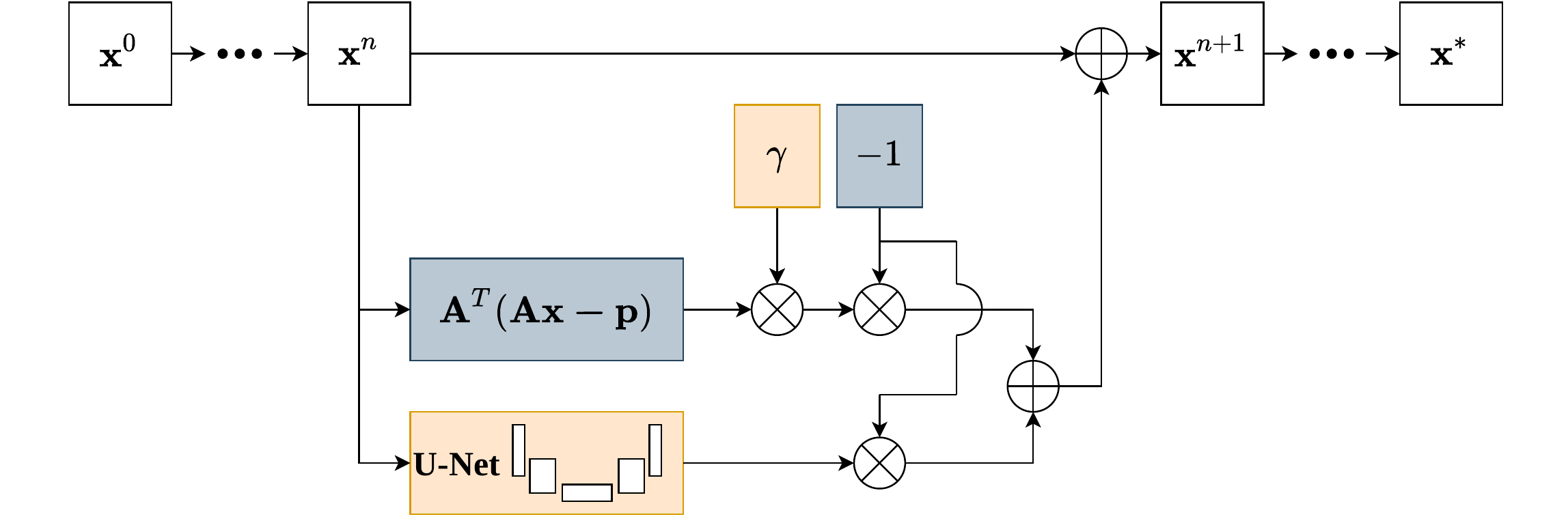}
        \label{fig:network_unrolled}
    }
    \hfil
    \subfloat[]{
        \includegraphics[height=2.9cm]{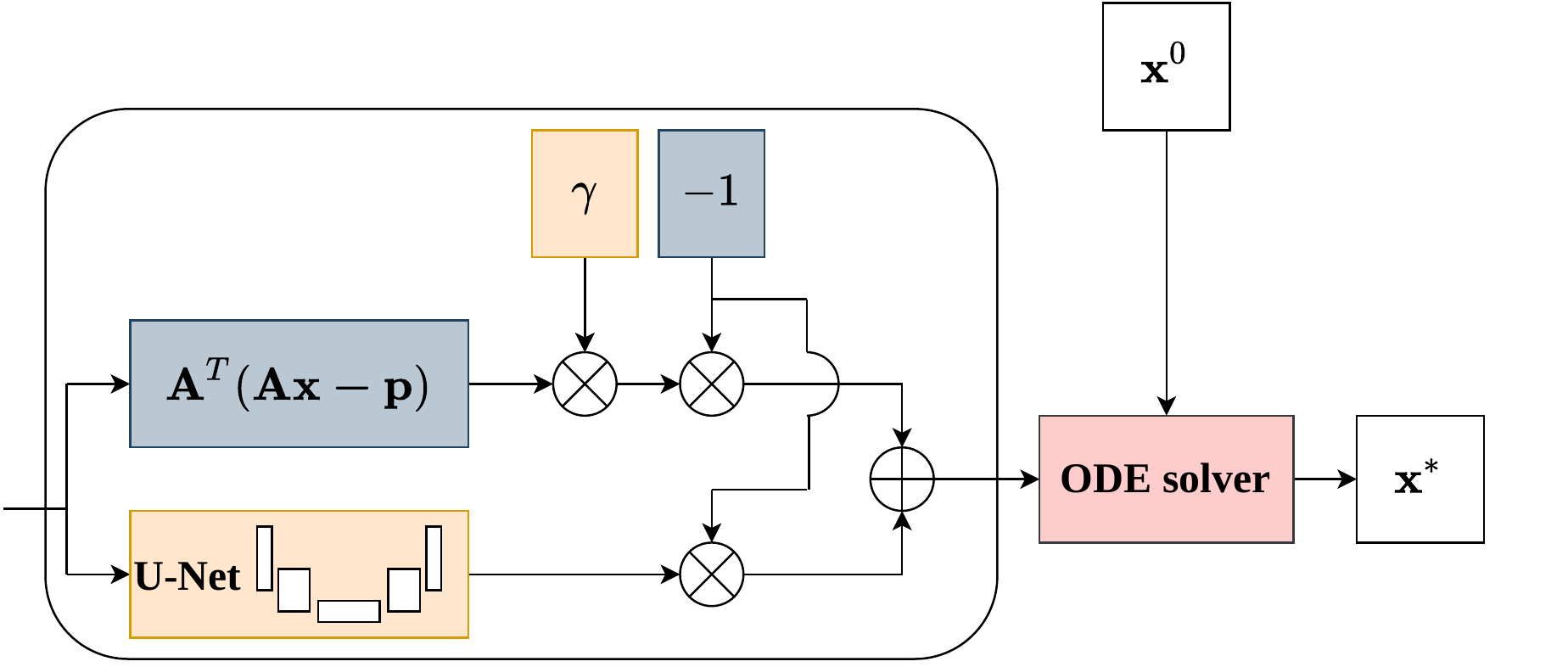}
        \label{fig:network_ode}
    }
    \caption{Comparison of an unrolled network (a) and a neural ODE-based version (b). While the unrolled version explicitly repeats the same network block for a fixed number of steps, the ODE solver receives only one network block parameterizing the temporal derivative of the volume. The solver computes the numerical integration internally. All trainable parts of our architecture are highlighted in yellow, the fixed parts are gray.}
\end{figure}

\subsection{Learned Iterative CT Reconstruction}
The forward model of a CT acquisition can be written as 
\begin{equation}
    \mathbf{p} = \mathbf{Ax} + \mathbf{\epsilon} \enspace ,
\end{equation}
where $\mathbf{x} \in \mathbb{R}^M$ is the volume, $\mathbf{p} \in \mathbb{R}^N$ is the projection data, $\mathbf{A} \in \mathbb{R}^{N \times M}$ is the forward operator defined by the imaging geometry (cone-beam in this case), and $\epsilon \in \mathbb{R}^N$ is additive noise. Typically, recovering the volume $\mathbf{x}$ from the measured data $\mathbf{p}$ is an ill-posed inverse problem meaning that $\mathbf{A}$ is not square and there are multiple solutions for $\mathbf{x}$ which are consistent with the measured data $\mathbf{p}$. Hence, the reconstruction is formulated as a regularized optimization problem
\begin{equation}
    \mathbf{x}^* = \argmin_{\mathbf{x}} \{D(\mathbf{x, p}) + \mu R(\mathbf{x})\} \enspace .
\end{equation}
Here, $D:\mathbb{R}^M \times \mathbb{R}^N \rightarrow \mathbb{R}$ is a function which measures the consistency of volume $\mathbf{x}$ and measured data $\mathbf{p}$. We choose the data consistency term in a least squares sense given as $D(\mathbf{x, p}) = \frac{1}{2} \|\mathbf{Ax - p}\|_2^2$. $R: \mathbb{R}^M \rightarrow \mathbb{R}$ is a regularizer which helps finding a favorable solution $\mathbf{x}^*$ and is weighted against the data consistency term by a scalar $\mu \in \mathbb{R}$. Using a simple gradient descent optimization scheme, the resulting update formula is
\begin{equation}
    \mathbf{x}^{n+1} = \mathbf{x}^n - \lambda \nabla \{D(\mathbf{x, p}) + \mu R(\mathbf{x})\} = \mathbf{x}^n - \lambda (\mathbf{A}^T(\mathbf{Ax - p}) + \mu \nabla R(\mathbf{x})) \enspace ,
\end{equation}
where $\mathbf{A}^T \in \mathbb{R}^{M \times N}$ is the adjoint operator of $\mathbf{A}$, $\lambda \in \mathbb{R}$ is a sufficiently small step size, and $n$ is the iteration index.
To learn a flexible regularizer from data, we replace its gradient by a network $N_{\theta}:\mathbb{R}^M \rightarrow \mathbb{R}^M$ with free parameters $\theta$ which allows to fit the regularizing component directly from data. The final update formula is given as 
\begin{equation}
\label{eq:update}
    \mathbf{x}^{n+1} = \mathbf{x}^n - \lambda (\mathbf{A}^T(\mathbf{Ax - p}) + \mu N_{\theta}(\mathbf{x})) \enspace .
\end{equation}
If $\mathbf{x}$ represents a 2D image, this equation could inspire an unrolled network architecture (\figurename~\ref{fig:network_unrolled}) and the parameters $\theta$ can be trained from pairs of input projection data and ground truth reconstruction for a fixed number of unrolled iterations \cite{chen2018learn}. This is infeasible for 3D cone-beam data due to extremely high GPU memory requirements. 

\subsection{Neural Ordinary Differential Equations}
Equation~\ref{eq:update} has a residual form of the type 
\begin{equation}
    \mathbf{x}^{n+1} = \mathbf{x}^n + f_{\theta}(\mathbf{x, p}) \enspace ,
\end{equation} with $f_{\theta}(\mathbf{x, p}) = - \lambda (\mathbf{A}^T(\mathbf{Ax - p}) + \mu N_{\theta}(\mathbf{x}))$. The function $f_{\theta}$ describes how the volume $\mathbf{x}^n$ changes incrementally. Chen et al. \cite{chen2018} proposed to regard such residual neural network architectures as the numerical integration of some underlying continuous ordinary differential equation. Here, the continuous differential equation would be $\frac{d\mathbf{x}}{dt} = f_{\theta}(\mathbf{x(t), p})$. Following that idea, a full unrolled iterative reconstruction is similar to the solution of an initial value problem of the given ODE starting from an initial condition $\mathbf{x}^0$ integrated until some end time $T$
\begin{equation}
\label{eq:integral}
    \mathbf{x}^* = \mathbf{x}^0 + \int_{0}^{T} f_{\theta}(\mathbf{x(t), p}) \mathrm{dt} \enspace .
\end{equation}
There exist a number of different numerical solvers to approximate solutions of such initial value problems. In this work, we use a fixed step-size Runge-Kutta solver of order 4 which integrates Eq.~\ref{eq:integral} by dividing the interval $[0, .., T]$ into a fixed number of steps $S$ to solve the integral numerically. This highlights the analogy to residual networks of depth $S$. 

To be able to combine this ODE-based problem formulation with a trainable network architecture, we need to compute a loss that is based on the output of the ODE solver and use its gradient to update the weights $\theta$ contained in $f_{\theta}$. As demonstrated in \cite{chen2018}, this gradient can be obtained without backpropagating through the internals of the solver. Instead, the solver is regarded as a black box and the gradient with respect to $\theta$ is computed by solving another ODE backward in time (adjoint sensitivity method). This allows to obtain gradients with a memory cost that is independent of the number of steps $S$ taken by the solver. Coming back to the analogy with residual networks, we can unroll the reconstruction problem in many steps using neural ODEs without further increasing the memory cost. 

\subsection{Network Architecture}
In the neural ODE setting, a neural network defines the dynamics of the system, i.e., its temporal derivative. Following the classical problem formulation in Eq.~\ref{eq:update}, we design this network using two branches: (1) A data consistency branch with no trainable parameters incorporating the system's forward and backward model as known operator and (2) a regularization branch which is trained from data. Figure~\ref{fig:network_ode} illustrates the proposed network architecture. The data consistency branch implements the term $\mathbf{A}^T(\mathbf{Ax - p})$. Operators $\mathbf{A}$ and $\mathbf{A}^T$ are the CT forward and backprojection under the correct cone-beam geometry, respectively. We use the differentiable version of these operators described in \cite{syben2019} which computes analytical gradients and allows for a direct embedding of these operators in neural networks. For the regularization branch, we use a standard 3D U-Net \cite{wolny2020} with depth $4$ and $8$ feature maps on the first level which are doubled in each stage, ReLU activation function, and instance normalization. In total, this leads to a network with \num{255000} free parameters. We further introduce an additional single trainable parameter $\gamma$ (referred to as data consistency weight) which is multiplied to the output of the data consistency branch before adding the output of both branches together in order to enable a data-optimal weighting between the data consistency and regularization component. This network together with the initialization $\mathbf{x}^0$ is passed to the ODE solver.  

\section{Experiments}

\subsection{Data}
The data set consists of $42$ walnuts scanned under cone-beam CT geometry (cone angle: $40^{\circ}$) \cite{der2019}. It contains the raw projection data and a corresponding ground truth reconstruction. The projection images are acquired on three circular trajectories on different heights along the walnuts' long axes with a full rotation of $360^{\circ}$ divided into \num{1200} angular steps each. The ground truth reconstruction is computed iteratively from the full set of acquired projections. 
As input to our algorithm, we use projection data from only the central one of the three trajectories and downsample the projections by a factor of $10$ in angular direction and $2$ in the spatial directions. This results in $120$ projection images of size $384 \times 486$ pixels with an angular increment of $3^{\circ}$ covering a full rotation of $360^{\circ}$. Hence, the algorithm has much less projection data available than has been used for computing the ground truth reconstruction which serves as learning target and is downsampled by a factor of $2$ resulting in volumes of size $251^3$. We use walnut number $1$ for validation and walnut number $2$ for testing. The rest is used for training. The used data, its preprocessing and the train-test-splits are identical to \cite{rudzusika2021}. 

\subsection{Training Details}
We use the neural ODE solver provided by \cite{chen2018}. Integration of Eq.~\ref{eq:integral} is performed from $0$ to $T=1$ with a fixed step size of $0.05$. This results in $20$ steps and $80$ evaluations of the network per forward and backward pass as the fourth order Runge-Kutta solver takes four evaluations per step. The initial volume $\mathbf{x}^0$ is a Feldmann-Davis-Kress (FDK) reconstruction of the projection data. The last layer of the U-Net is initialized with zeros such that the regularizer has no influence upon initialization and the data consistency weight is initialized with $\gamma=0.01$. The parameters are optimized by an Adam optimizer with learning rate \num{1e-4} for the U-Net and \num{1e-2} for the data consistency weight. Training is performed with batch size 1 for $80$ epochs and an L1-Loss evaluated only inside the cylindrical scan field of view (FOV) captured in each projection. The model weights corresponding to the lowest validation loss during training are selected for further evaluation. 

\subsection{Reference Methods}
The reconstruction of the proposed method is compared to (1) an FDK reconstruction, (2) a SIRT reconstruction with non-negativity constraint ($500$ iterations) \cite{vanaarle2016} and (3) a total variation (TV) regularized reconstruction using gradient updates ($300$ iterations) \cite{syben2019}.
\begin{figure}[!t]
    \centering
    \includegraphics[width=0.505\textwidth,trim={0.2cm 0.3cm 0 0.3cm},clip]{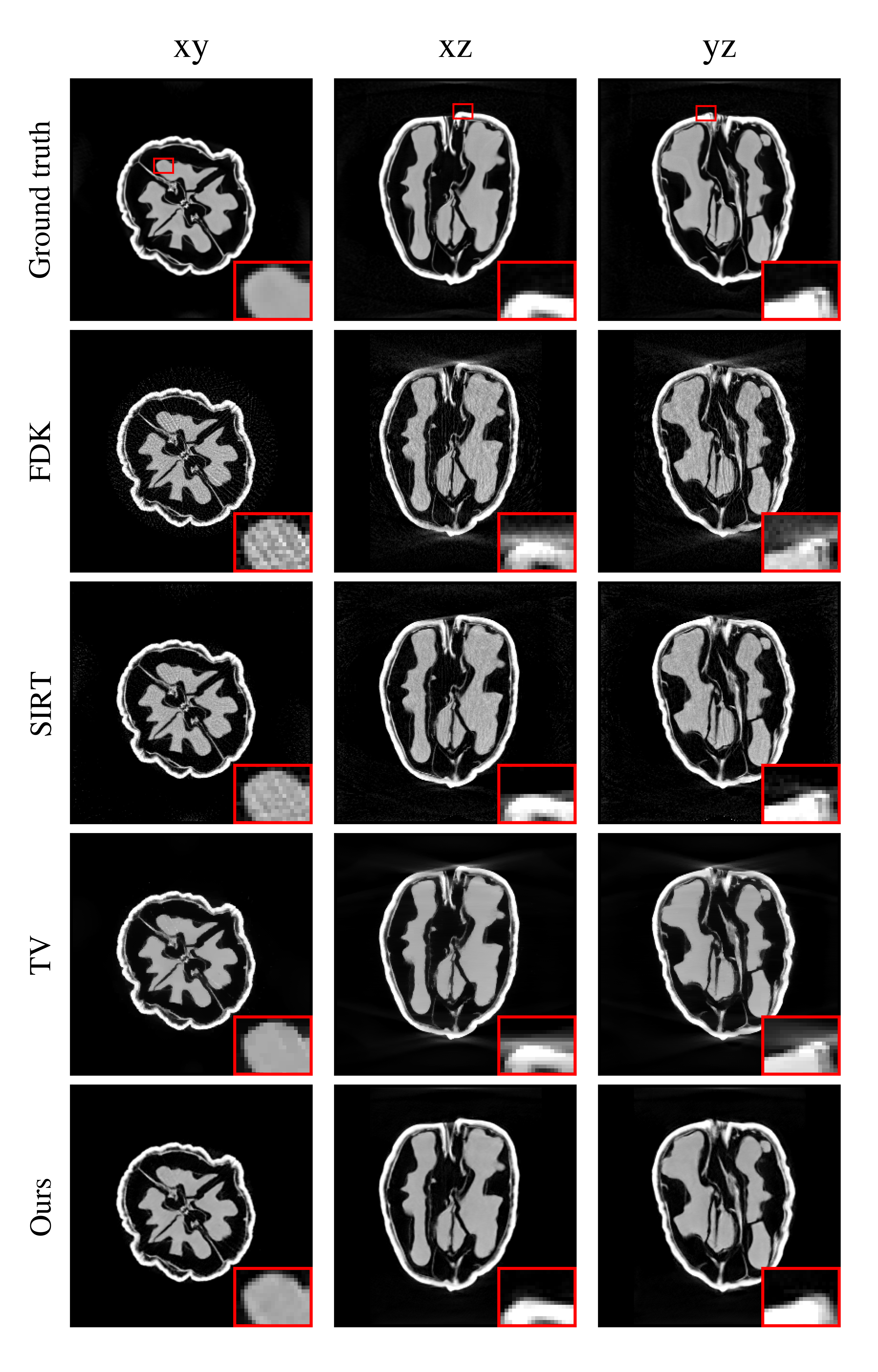}
    \caption{Reconstructed center slices of the test walnut along each dimension using an analytical FDK algorithm, two classical iterative algorithms (SIRT, TV) and the proposed method. The gray value window is $0$ to \SI{0.06}{\milli\meter^{-1}}. Zoomed regions are indicated in red.}
    \label{fig:qualitative}
\end{figure}

\section{Results}
The proposed model requires a maximum of \SI{13}{GB} GPU memory during training. Reconstruction results of the test walnut are shown in \figurename~\ref{fig:qualitative}. The FDK reconstruction exhibits strong streaking artifacts in the xy-plane as well as cone-beam artifacts at the top and bottom of the walnut visible in the xz- and yz-planes. The iterative SIRT algorithm only partly removes these artifacts. The TV regularized algorithm produces a very smooth image with homogeneous gray values for different structures inside the walnut. Nevertheless, cone-beam artifacts in the xz- and yz-plane are still visible. In contrast, our method achieves images with no noticeable cone-beam artifacts. Additionally, it performs well in removing the streaks in the xy-plane and produces images which are visually closest to the ground truth. The proposed method also performs best regarding all quantitative metrics (Tab.~\ref{tab:quantitative}). Compared to the SIRT which is the second-best performing method, the RMSE is reduced by \SI{31.7}{\%} and PSNR and SSIM are increased by \SI{11.1}{\%} and \SI{11.3}{\%}, respectively. All metrics have only been evaluated inside the cylindrical scan FOV captured in each projection. Concerning reconstruction time, our method lies between the two investigated iterative methods with a run time of \SI{43.4}{\s}. The data consistency weight converges to a value of $\gamma=0.036$. 
\begin{table}[!t]
\renewcommand{\arraystretch}{1.3}
\caption{Quantitative results in terms of root mean squared error (RMSE), peak signal-to-noise ratio (PSNR), and structural similarity index measure (SSIM).}
\label{tab:quantitative}
\centering
\begin{tabular}{ @{} l l l l l @{} }
\hline
                                & FDK      & SIRT      & TV        & Ours \\ 
\hline
RMSE [$\cdot 10^{-3}$] $\downarrow$  & 3.735    & 2.321     & 2.688     & 1.586 \\  
PSNR $\uparrow$                 & 25.682   & 29.813    & 28.539    & 33.121 \\
SSIM $\uparrow$                 & 0.562    & 0.813     & 0.777     & 0.904 \\
\hline
\end{tabular}
\end{table}

\section{Discussion}
Our proposed method is able to train a network inspired from classical iterative reconstruction for 3D cone-beam data incorporating a trainable regularizer with a GPU memory consumption which is independent of the number of incremental update steps on the volume. We can reconstruct volumes of practically relevant size ($251^3$) while using only \SI{13}{GB} of GPU memory during training. This is feasible with a single recent consumer graphics card.

The considered reconstruction problem is severely ill-posed due to the strong undersampling of projection data in angular direction. Hence, the analytical FDK reconstruction leads to strong artifacts in the reconstructed images. The trained algorithm removes the noise and streak artifacts successfully. While the classical TV-regularized iterative reconstruction also performs well in this regard, the proposed algorithm is the only one which is able to remove the cone-beam artifacts. We hypothesize that one main advantage of the learned regularizer parameterized by the U-Net over hand-crafted ones such as TV is its larger receptive field. It can suppress artifacts with non-local extent such as the cone-beam artifacts while TV depends only on the local gradient information in the image. A detailed comparison of our method to other learning-based approaches, such as \cite{rudzusika2021}, will be performed in future work.

Once trained, the reconstruction time of the presented method is comparable to that of classical iterative algorithms. 
Training time is rather high due to the high number of network evaluations. Potentially, using an adaptive ODE solver instead of the fixed step size solver can shorten training and inference times by adaptively adjusting the step size and hence the number of network evaluations to a given tolerance. 

\section{Conclusion}
This paper presents a method which uses neural ODEs to train a cone-beam reconstruction algorithm inspired by iterative reconstruction schemes. It ensures consistency with the measured data by incorporating a data consistency branch which exploits analytical knowledge about the physical operator connecting sinogram and image domain along with a trained regularizer. The proposed method outperforms well-known FDK and iterative reconstruction algorithms on the used walnut data set and is able to remove artifacts with non-local extent such as the cone-beam artifacts while being tractable concerning GPU memory.  

\section*{Acknowledgment}
The research leading to these results has received funding from the European Research Council (ERC) under the European Union’s Horizon 2020 research and innovation program (ERC Grant No. 810316).

\bibliographystyle{unsrt}  
\bibliography{main}  

\begin{thebibliography}{10}

\bibitem{chen2018learn}
Hu~Chen, Yi~Zhang, Yunjin Chen, Junfeng Zhang, Weihua Zhang, Huaiqiang Sun,
  Yang Lv, Peixi Liao, Jiliu Zhou, and Ge~Wang.
\newblock {LEARN: Learned experts’ assessment-based reconstruction network
  for sparse-data CT}.
\newblock {\em IEEE Trans. Med. Imag.}, 37(6):1333--1347, 2018.

\bibitem{jin2017}
Kyong~Hwan Jin, Michael~T. McCann, Emmanuel Froustey, and Michael Unser.
\newblock {Deep Convolutional Neural Network for Inverse Problems in Imaging}.
\newblock {\em IEEE Trans. Image Process.}, 26(9):4509--4522, 2017.

\bibitem{adler2021}
Jonas Adler.
\newblock {Learned Iterative Reconstruction}.
\newblock {\em Handbook of Mathematical Models and Algorithms in Computer
  Vision and Imaging: Mathematical Imaging and Vision}, pages 1--22, 2021.

\bibitem{hammernik2018}
Kerstin Hammernik, Teresa Klatzer, Erich Kobler, Michael~P Recht, Daniel~K
  Sodickson, Thomas Pock, and Florian Knoll.
\newblock Learning a variational network for reconstruction of accelerated
  {MRI} data.
\newblock {\em Magn. Reson. Med.}, 79(6):3055--3071, 2018.

\bibitem{gadjimuradov2021}
Fasil Gadjimuradov, Thomas Benkert, Marcel~Dominik Nickel, and Andreas Maier.
\newblock {Robust partial Fourier reconstruction for diffusion-weighted imaging
  using a recurrent convolutional neural network}.
\newblock {\em Magn. Reson. Med.}, 2021.

\bibitem{vishnevski2019}
Valery Vishnevskiy, Richard Rau, and Orcun Goksel.
\newblock {Deep Variational Networks with Exponential Weighting for Learning
  Computed Tomography}.
\newblock In {\em Proc. MICCAI}, pages 310--318. Springer International
  Publishing, 2019.

\bibitem{adler2018}
Jonas Adler and Ozan Öktem.
\newblock {Learned Primal-Dual Reconstruction}.
\newblock {\em IEEE Trans. Med. Imag.}, 37(6):1322--1332, 2018.

\bibitem{wu2019}
Dufan Wu, Kyungsang Kim, and Quanzheng Li.
\newblock Computationally efficient deep neural network for computed tomography
  image reconstruction.
\newblock {\em Med. Phys.}, 46(11):4763--4776, 2019.

\bibitem{hauptmann2020}
Andreas Hauptmann, Jonas Adler, Simon Arridge, and Ozan {\"O}ktem.
\newblock Multi-scale learned iterative reconstruction.
\newblock {\em IEEE Trans. Comput. Imag.}, 6:843--856, 2020.

\bibitem{rudzusika2021}
Jevgenija Rudzusika, Buda Bajic, Ozan {\"O}ktem, Carola-Bibiane Sch{\"o}nlieb,
  and Christian Etmann.
\newblock {Invertible Learned Primal-Dual}.
\newblock In {\em Proc. NeurIPS}, 2021.

\bibitem{kellman2020}
Michael Kellman, Kevin Zhang, Eric Markley, Jon Tamir, Emrah Bostan, Michael
  Lustig, and Laura Waller.
\newblock Memory-efficient learning for large-scale computational imaging.
\newblock {\em IEEE Trans. Comput. Imag.}, 6:1403--1414, 2020.

\bibitem{chen2018}
Ricky~TQ Chen, Yulia Rubanova, Jesse Bettencourt, and David Duvenaud.
\newblock Neural ordinary differential equations.
\newblock In {\em Proc. NeurIPS}, pages 6572--6583, 2018.

\bibitem{chen2020}
Eric~Z Chen, Terrence Chen, and Shanhui Sun.
\newblock {MRI Image Reconstruction via Learning Optimization Using Neural
  ODEs}.
\newblock In {\em Proc. MICCAI}, pages 83--93. Springer, 2020.

\bibitem{syben2019}
Christopher Syben, Markus Michen, Bernhard Stimpel, Stephan Seitz, Stefan
  Ploner, and Andreas~K Maier.
\newblock {PYRO-NN: Python reconstruction operators in neural networks}.
\newblock {\em Med. Phys.}, 46(11):5110--5115, 2019.

\bibitem{wolny2020}
Adrian Wolny, Lorenzo Cerrone, Athul Vijayan, et~al.
\newblock {Accurate and versatile 3D segmentation of plant tissues at cellular
  resolution}.
\newblock {\em eLife}, 9:e57613, 2020.

\bibitem{der2019}
Henri Der~Sarkissian, Felix Lucka, Maureen van Eijnatten, Giulia Colacicco,
  Sophia~Bethany Coban, and Kees~Joost Batenburg.
\newblock {A cone-beam X-ray computed tomography data collection designed for
  machine learning}.
\newblock {\em Sci. Data}, 6(1):1--8, 2019.

\bibitem{vanaarle2016}
Wim van Aarle, Willem~Jan Palenstijn, Jeroen Cant, Eline Janssens, Folkert
  Bleichrodt, Andrei Dabravolski, Jan~De Beenhouwer, K.~Joost Batenburg, and
  Jan Sijbers.
\newblock {Fast and flexible X-ray tomography using the ASTRA toolbox}.
\newblock {\em Opt. Express}, 24(22):25129--25147, Oct 2016.

\end{thebibliography}

\end{document}